\documentclass[12pt,letterpaper]{article}

\usepackage{epsfig}
\usepackage{amsmath}
\usepackage{amsfonts}
\usepackage{amssymb}
\usepackage{graphicx}

\begin{document}

\begin{flushright}
{\tt BROWN-HET-1629} \\ {\tt WITS-CTP-096 \quad}
\end{flushright}

\ \\

\begin{center}

{\Huge {\bf $\mathbf{S=1}$ in $\mathbf{O(N)}$/HS duality}}

\vspace{0.8cm}

{Robert de Mello Koch$^a$, Antal Jevicki$^b$, Kewang Jin$^c$, \\
Jo\~{a}o P. Rodrigues$^a$, and Qibin Ye$^b$}

\vspace{1cm}

{\it $^a$National Institute for Theoretical Physics, \\
School of Physics and Centre for Theoretical Physics, \\ 
University of the Witwatersrand, Wits 2050, South Africa \\
\tt{\small robert@neo.phys.wits.ac.za}, \tt{\small joao.rodrigues@wits.ac.za}
}

\vspace{0.5cm}

{\it $^b$Department of Physics, Brown University, Providence, RI 02912, USA \\
\tt{\small antal\_jevicki@brown.edu}, \tt{\small qibin\_ye@brown.edu}
}

\vspace{0.5cm}

{\it $^c$Institut f\"{u}r Theoretische Physik, ETH-Z\"{u}rich, CH-8093 Z\"{u}rich, Switzerland \\
\tt{\small jinke@itp.phys.ethz.ch}
}

\vspace{1.0cm}

\begin{abstract}

Following the work of Maldacena and Zhiboedov, we study the implementation of the Coleman-Mandula theorem in the free $O(N)$/Higher Spin correspondence. In the bi-local framework we first define an $S$-matrix for scattering of collective dipoles. Its evaluation in the case of free UV fixed point theory leads to the result $S=1$ stated in the title. We also present an appropriate field transformation that is seen to transform away all the non-linear $1/N$ interactions of this theory.  A change of boundary conditions and/or external potentials results in a nontrivial $S$-matrix.

\end{abstract}

\end{center}

\thispagestyle{empty}

\pagebreak

\setcounter{page}{1}

\section{Introduction}

The duality involving $O(N)$ vector models and Vasiliev's Higher Spin Gravity in Anti de Sitter space represent some of the simplest examples of AdS/CFT and as such is a topic of vigorous recent studies \cite{Klebanov:2002ja, Sezgin:2002rt, Das:2003vw, Giombi:2009wh, Giombi:2010vg, Koch:2010cy, Douglas:2010rc, Jevicki:2011ss, Shenker:2011zf, Maldacena:2011jn, Bekaert:2012ux, Vasiliev:2012vf, Maldacena:2012sf}. Equally interesting is the correspondence between 2d minimal CFT's and 3d Chern-Simons Higher Spin Gravity \cite{Henneaux:2010xg, Campoleoni:2010zq, Gaberdiel:2010pz, Gaberdiel:2011zw, Chang:2011mz, Gary:2012ms, Gaberdiel:2012ku}. These large $N$ dualities involve quantum field theories that have been thought to be understood for some time and a relatively novel version of HS Gravity built on a single Regge trajectory. These theories feature many properties that have been unreachable in String Theory, in particular the structure and explicit form of the higher spin gauge symmetry group. They also offer a potentially solvable framework for studies of black hole formation and de Sitter theory itself \cite{Anninos:2011ui, Ng:2012xp}.

In the case of three dimensional $O(N)$ vector field theory, one has two conformally invariant fixed points, the UV and the IR one. The HS duals are given by the same Vasiliev theory \cite{Vasiliev:1995dn, Vasiliev:2003ev, Bekaert:2005vh, Metsaev:2005ar, Metsaev:2007rn, Polyakov:2010sk, Iazeolla:2011cb} but with different boundary conditions on the scalar field \cite{Klebanov:2002ja}. This provides a simple relationship between a (HS) theory dual to a free $N$-component scalar field (UV) and the nontrivial dual corresponding to the IR CFT. The correspondence provided by the free $O(N)$ scalar field theory is then of central interest. This theory is characterized by an infinite sequence of conserved currents which are themselves boundary duals of higher spin fields and whose correlation functions represent a point of comparison \cite{Giombi:2009wh, Giombi:2010vg} between the two descriptions.

Since conserved currents imply the existence of an infinite sequence of conserved charges and higher symmetries, one is faced with the question regarding the implementation (and implication) of the Coleman-Mandula theorem. This question was raised and addressed in the recent work of Maldacena and Zhiboedov \cite{Maldacena:2011jn, Maldacena:2012sf} who were able to show the existence of conserved currents (charges) implies that the correlation functions are built in terms of free fields. This demonstrates the simplicity of the corresponding Vasiliev theory. One still, however, has the question regarding the triviality of the theory in the bulk. In standard field theories this question is addressed (and answered) through the $S$-matrix. The Coleman-Mandula theorem in particular would imply $S=1$ for theories with higher symmetries. Due to the equivalence theorem (under field transformations) this means that there exists a field redefinition which linearizes the field equations. In the AdS/CFT framework one sometimes thinks of the correlators as taking the role analogous to an $S$-matrix. A proposal along this line offered by Mack \cite{Mack:2009mi, Mack:2009gy} has been nicely implemented in recent works \cite{Penedones:2010ue, Fitzpatrick:2011ia, Paulos:2011ie, Nandan:2011wc}. If this analogy is taken at face value, one has the puzzling fact that this $S$-matrix is non-trivial, even for the correspondence based on free theory.

We have in \cite{Das:2003vw,Koch:2010cy} formulated a constructive approach to bulk AdS duality and HS Gravity in terms of bi-locals. It leads to a nonlinear, interacting theory (with $1/N$ as the coupling constant) which was seen to posses all the properties of the dual AdS theory. This theory reproduces arbitrary-point correlation functions and provides a construction of HS theory (in various gauges) based on CFT \cite{Jevicki:2011ss}. The construction also, as we will explain, offers a framework for defining and calculating an $S$-matrix and addressing the implementation of the Coleman-Mandula theorem in the nonlinear bulk framework. The construction is based on the two-particle collective dipole and its interactions in the large $N$ limit. It has been known since the early works that nontrivial collective phenomena can appear ``even'' for free theories (for example excitations near the large $N$ fermion surface). In the present case we are led to consider the $S$-matrix for collective dipoles: the corresponding LSZ reduction formula is easily stated as a limit of bi-local correlators. Its evaluation will produce the result $S=1$ as claimed in the title.\footnote{We mention that this is analogous to an earlier situation involving the $c=1$ matrix model with 2d string correspondence where one had the statement ``$S=1$ for $c=1$"  demonstrated in \cite{Gross:1991qp}. The only difference is that the collective boson (representing fluctuations above the fermion surface) is now replaced by the collective dipole.}

$S=1$ implies triviality, namely that interactions can be removed by a nonlinear transformation of fields (by this we mean the $1/N$ interactions which equal $G_N$ interactions in Vasiliev's theory). We demonstrate this for the nonlinear dipole representation, where we establish a construction of a nonlinear field transformation that linearizes the bi-local field theory. One might (naively) wonder if the linearization is (not) just a change from bulk HS fields to the $N$-component CFT fields. One should remember however that the large $N$ duality is Quantum $\leftrightarrow$ Classical, namely that for recovering the classical AdS theory quantum effects in CFT need to be taken into account. Consequently the issue of a transformation between the two boils down to the construction of the (elusive) master field construction at large $N$. Some knowledge on this exists \cite{Gopakumar:1994iq, Douglas:1994kw}.

The content of this paper is as follows. In section \ref{sec:CFT} we summarize the basics of the $O(N)$ model and the associated (nonlinear) bi-local field construction of \cite{Koch:2010cy, Jevicki:2011ss}. In section \ref{sec:CM}, after a summary of the Maldacena-Zhiboedov result, we discuss the differences between ``Boundary $S$-matrix'' and ``Collective $S$-matrix'' that we propose. In particular we give an LSZ formula for the $S$-matrix and evaluate the associated three- and four-point amplitudes using the cubic and quartic vertices of the $1/N$ theory demonstrating the result $S=1$. In section \ref{sec:Field} we present a construction of a nonlinear bi-local field transformation that linearizes the theory. Conclusions are given in section \ref{sec:Con}.

\section{Bi-local representation of $\mathbf{O(N)}$ CFT$_\mathbf{3}$} \label{sec:CFT}

In the $O(N)$/Higher Spin duality one starts with a $N$-component scalar field theory
\begin{equation}
\mathcal{L}= \frac{1}{2}\partial_\mu \phi^a \partial^\mu \phi^a+\frac{g}{4}(\phi\cdot\phi)^2 \ , \qquad a=1,\cdots,N
\end{equation}
where $\phi^a=\phi^a(t,\vec{x})=\phi^a(x^+,x^-,x^\perp)$. This theory features two critical points with conformal symmetry: the UV fixed point at zero coupling ($g=0$) and the nontrivial IR fixed point at nonzero value of the constant coupling constant. The latter can be evaluated in the large $N$ limit and serves as the classic example of critical phenomena in 3d.

For the correspondence with higher spin fields, a central role is played by the sequence of traceless and symmetric higher spin currents
\begin{equation}
J_{\mu_1 \mu_2 \cdots \mu_s}=\sum_{k=0}^s (-1)^k \begin{pmatrix} s-1/2 \cr k \end{pmatrix}
\begin{pmatrix} s-1/2 \cr s-k \end{pmatrix} \partial_{\mu_1} \cdots \partial_{\mu_k} \phi^a \;
\partial_{\mu_{k+1}} \cdots \partial_{\mu_s} \phi^a -\text{traces}
\end{equation} 
which are exactly conserved in the free case. These operators can be summarized in the semi bi-local form by the generating function
\begin{equation}
\mathcal{O}(x,\epsilon)=\phi^a(x-\epsilon)\sum_{n=0}^\infty \frac{1}{(2n)!}
\left(2\epsilon^2 \overleftarrow{\partial_x} \cdot \overrightarrow{\partial_x}
-4(\epsilon \cdot \overleftarrow{\partial_x})(\epsilon \cdot \overrightarrow{\partial_x})\right)^n
\phi^a(x+\epsilon)
\end{equation}
where $\epsilon^2=0$ (in order to satisfy the traceless condition). As a result, $\epsilon$ represents a cone with a two dimensional coordinate and altogether $\mathcal{O}(x,\epsilon)$ is a five dimensional semi bi-local field. The currents that it generates represent boundary duals of AdS$_4$ higher spin fields
\begin{equation}
J_{\mu_1 \mu_2 \cdots \mu_s}(x) \leftrightarrow \mathcal{H}_{\hat{\mu}_1 \hat{\mu}_2 \cdots \hat{\mu}_s}(x,z\rightarrow 0)
\end{equation}
where $ds^2=\frac{dx^2+dz^2}{z^2}$ is the AdS$_4$ metric. In the AdS/CFT correspondence, correlation functions of currents are to match up with the boundary transition amplitude (sometimes referred to as the boundary $S$-matrix) of the higher dimensional AdS theory. A successful demonstration of this was accomplished in the three-point case by Giombi and Yin \cite{Giombi:2009wh, Giombi:2010vg} who were able to match the two critical points of the vector model with two versions of Vasiliev's Higher Spin Gravity in AdS$_4$. The trivial and nontrivial fixed points are seen as conjectured by Klebanov and Polyakov \cite{Klebanov:2002ja} to correspond to different boundary conditions involving the lowest spin ($s=0$) field.

A constructive approach for this AdS$_4$/CFT$_3$ correspondence was given in \cite{Das:2003vw}, based on the notion of collective fields. These are given by bi-local invariants of the $O(N)$ field theory
\begin{equation}
\Phi(x,y)\equiv \phi(x)\cdot\phi(y)=\sum_{a=1}^N\phi^a(x)\cdot\phi^a(y) 
\end{equation}
and close under the Schwinger-Dyson equations (in the large $N$ limit). These operators represent a more general set than the conformal fields $\mathcal{O}(x,\epsilon)$ since there is no restriction to a cone. The collective action evaluates the complete $O(N)$ invariant partition function
\begin{equation}
Z=\int [d\phi^a(x)] e^{-S[\phi]}=\int \prod_{x,y} [d\Phi(x,y)] \mu(\Phi) e^{-S_c[\Phi]}
\end{equation}
where the measure is given by $\mu(\Phi)=(\det \Phi)^{V_x V_p}$ with $V_x=L^3$ the volume of space and $V_p=\Lambda^3$ the volume of momentum space where $\Lambda$ is the momentum cutoff. Explicitly one has the collective action
\begin{equation}
S_c[\Phi]=\text{Tr}\left[-(\partial_x^2+\partial_y^2)\Phi(x,y)+V \right]+\frac{N}{2}\text{Tr} \, \ln \Phi(x,y)
\end{equation}
where the trace is defined as ${\rm Tr} \, B=\int d^3 x \, B(x,x)$. This collective action is nonlinear, with $1/N$ appearing as the expansion parameter. Through the identification of $1/N $ with $G_N$ (the coupling constant of higher spin gravity), this collective field representation provides a bulk description of the dual AdS theory. One also has a natural (star) product defined as $(\Psi \star \Phi)(x,y)=\int dz \, \Psi(x,z)\Phi(z,y)$. 

\begin{figure}
\begin{center}
\includegraphics[width=0.55\textwidth]{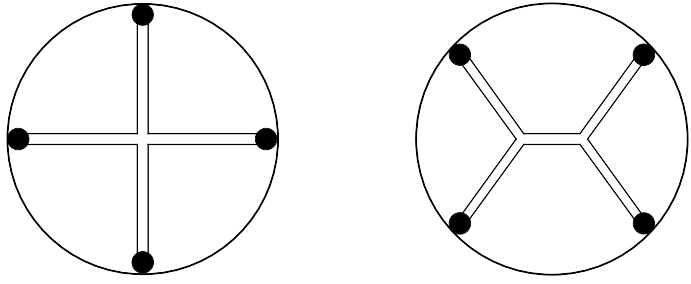}  
\caption{Illustration of the four-point collective field diagrams.}
\label{fourdiagram}
\end{center}
\end{figure}

The perturbation expansion is defined in this (bi-local) space. The nonlinear equation of motion specified by $S_c$ gives the background in the expansion: $\Phi=\Phi_0+\frac{1}{\sqrt{N}}\eta$. Expanding about the background gives rise to an infinite number of interaction vertices \cite{deMelloKoch:1996mj} 
\begin{eqnarray}
S_c=S[\Phi_0]+\text{Tr}[\Phi_0^{-1} \eta \Phi_0^{-1} \eta]+\frac{g}{4}\eta^2+\sum_{n \ge 3} N^{1-n/2} \, \text{Tr}B^n \ , \quad \text{with} \quad B=\Phi_0^{-1} \eta \ .
\end{eqnarray}
The nonlinearities built into $S_c$ are precisely such that all invariant correlators: $\langle \phi(x_1) \cdot \phi(y_1) \cdots \phi(x_n) \cdot \phi(y_n)\rangle$ are now reproduced through the Witten (Feynman) diagrams with $1/N$ vertices (the four-point example is shown in Figure \ref{fourdiagram}). We stress that this nonlinear structure is there for both the free and the interacting fixed point.

This bi-local theory is expected to represent a (covariant) gauge fixing of Vasiliev's gauge invariant theory. An attempt at a gauge invariant formalism is given in \cite{Douglas:2010rc}. A large number of degrees of freedom are removed in fixing a gauge in any gauge theory and this simplification happens in Higher Spin Gravity too. A one-to-one relationship between bi-local and AdS higher spin fields can be demonstrated in a physical (single-time) picture. The existence of such a gauge and the discussion of the collective dipole underlying the collective construction is given in \cite{Jevicki:2011ss}.

The single-time formulation now involves the equal time bi-local
\begin{equation}
\Psi(t,\vec{x},\vec{y})=\sum_a \phi^a(t,\vec{x})\phi^a(t,\vec{y}) \label{bilocal}
\end{equation}
and its conjugate momenta: $\Pi(\vec{x},\vec{y})=-i\frac{\delta}{\delta\Psi(\vec{x},\vec{y})}$ with a Hamiltonian of the form
\begin{eqnarray}
H=2\text{Tr}(\Pi \Psi \Pi)+\frac{1}{2}\int d\vec{x} [-\nabla_{\vec{x}}^2\Psi(\tilde{x},\tilde{y})\vert_{\tilde{x}=\tilde{y}}]+\frac{N^2}{8}\text{Tr} \, \Psi^{-1} \ ,
\end{eqnarray}
where we have set the coupling constant $g=0$. This Hamiltonian again has a natural $1/N$ expansion, after a background shift
\begin{equation}
\Psi=\Psi_0+\frac{1}{\sqrt{N}} \eta \ , \qquad \Pi=\sqrt{N} \pi \ ,
\end{equation}
one gets a quadratic Hamiltonian
\begin{equation}
H^{(2)}=2 {\rm Tr}( \pi \Psi_0 \pi )+\frac{1}{8} {\rm Tr} (\Psi_0^{-1} \eta \Psi_0^{-1} \eta \Psi_0^{-1}) \ .
\end{equation}
Fourier transform the background field as well as the fluctuations
\begin{eqnarray}
\Psi^0_{\vec{x}\vec{y}} &=& \int d\vec{k} e^{i \vec{k} \cdot (\vec{x}-\vec{y})} \psi^0_{\vec{k}} \ , \qquad \psi^0_{\vec{k}}=\frac{1}{2 \sqrt{\vec{k}^2}} \ , \\
\eta_{\vec{x}\vec{y}} &=& \int d \vec{k}_1 d \vec{k}_2 e^{-i \vec{k}_1 \cdot \vec{x}+i \vec{k}_2 \cdot \vec{y}} \eta_{\vec{k}_1 \vec{k}_2} \ , \\
\pi_{\vec{x}\vec{y}} &=&  \int d \vec{k}_1 d \vec{k}_2 e^{+i \vec{k}_1 \cdot \vec{x}-i \vec{k}_2 \cdot \vec{y}} \pi_{\vec{k}_1 \vec{k}_2} \ ,
\end{eqnarray}
one finds the quadratic Hamiltonian in momentum space
\begin{equation}
H^{(2)}=\frac{1}{2} \int d \vec{k}_1 d \vec{k}_2 \; \pi_{\vec{k}_1 \vec{k}_2} \pi_{\vec{k}_1 \vec{k}_2}+\frac{1}{8} \int d \vec{k}_1 d \vec{k}_2 \; \eta_{\vec{k}_1 \vec{k}_2}
\left( \psi_{\vec{k}_1}^{0\,\,-1}+\psi_{\vec{k}_2}^{0\,\,-1} \right)^2\eta_{\vec{k}_1 \vec{k}_2}   
\end{equation}
representing the (singlet) spectrum $\omega_{\vec{k}_1 \vec{k}_2}=\sqrt{\vec{k}_1^2}+\sqrt{\vec{k}_2^2}$ of the $O(N)$ theory. A sequence of $1/N$ vertices representing interactions can be found similarly and the cubic and quartic interactions are given explicitly as
\begin{eqnarray}
H^{(3)}&=&\frac{2}{\sqrt{N}}\text{Tr}(\pi \eta \pi)-\frac{1}{8\sqrt{N}}\text{Tr} \Psi_0^{-1} \eta \Psi_0^{-1} \eta \Psi_0^{-1} \eta \Psi_0^{-1} \ , \label{cubic} \\
H^{(4)}&=&\frac{1}{8N}\text{Tr} \Psi_0^{-1} \eta \Psi_0^{-1} \eta \Psi_0^{-1} \eta \Psi_0^{-1} \eta \Psi_0^{-1} \ .
\label{quartic}
\end{eqnarray}
We note that the form of these vertices is the same for both the free (UV) and the interacting (IR) conformal theories (the only difference is induced by the different background shifts in these two cases).

One also has a null-plane version of this construction which would correspond to light-cone gauge higher spin theory.  This was used in \cite{Koch:2010cy} to demonstrate the one-to-one map between the two descriptions: the null-plane bi-locals $\Psi(x^+;x_1^-,x_2^-;x_1,x_2)$ and the higher spin fields $\mathcal{H}(x^+;x^-,x,z;\theta)$ in AdS$_4$, where $\theta$ is higher spin coordinate generating a sequence of higher spins. Both fields have same number of dimensions $1+2+2=1+3+1$, the same representation of the conformal group, and the same number of degrees of freedom.

The explicit canonical transformation was given in \cite{Koch:2010cy} as
\begin{eqnarray}
x^-&=&\frac{x_1^-p_1^++x_2^-p_2^+}{p_1^++p_2^+} \ , \\
x&=&\frac{x_1 p_1^++x_2 p_2^+}{p_1^++p_2^+} \ , \\
z&=&\frac{\sqrt{p_1^+ p_2^+}}{p_1^++p_2^+}(x_1-x_2) \ , \\
\theta&=&2\arctan \sqrt{p_2^+ / p_1^+} \ ,
\end{eqnarray}
where $p_i^+$ are the conjugate momenta of $x_i^-$. The map going from the bi-local field to the higher spin field is given by an integral transformation
\begin{eqnarray}
\text{$\mathcal{H}$}(x^-,x,z,\theta)&=&\int dp^+dp^xdp^z \, e^{i(x^-p^++xp^x+zp^z)}\int dp_1^+dp_1dp_2^+dp_2\cr
&&\delta(p_1^++p_2^+-p^+)\delta(p_1+p_2-p^x)\delta(p_1\sqrt{p_2^+/p_1^+}-p_2\sqrt{p_1^+/p_2^+}-p^z)\cr
&&\delta(2\arctan \sqrt{p_2^+/p_1^+}-\theta)\tilde{\Psi}(p_1^+,p_2^+,p_1,p_2) \ ,
\end{eqnarray}
where $\tilde{\Psi}(p_1^+,p_2^+,p_1,p_2)$ is the Fourier transform of the bi-local field $\Psi(x_1^-,x_2^-,x_1,x_2)$.

It was shown in \cite{Koch:2010cy} that under this transformation all the generators of collective field theory map into the generators of light-cone gauge Higher Spin Gravity in the form given by Metsaev \cite{Metsaev:1999ui}. In particular the quadratic bi-local Hamiltonian
\begin{equation}
P^{-(2)}=\int dx_1^- dx_1 dx_2^- dx_2 \, \Psi^\dagger \left(-\frac{\nabla_1^2}{2\partial_{x_1^-}}-\frac{\nabla_2^2}{2\partial_{x_2^-}} \right) \Psi
\end{equation}
takes an AdS$_4$ form
\begin{equation}
P^{-(2)}=\int dx^- dx dz d\theta \; \mathcal{H}^\dagger \left(-\frac{\partial_x^2+\partial_z^2}{2\partial_{x^-}} \right) \mathcal{H} \ .
\end{equation}

This establishes at the quadratic level the bi-local representation is identical to the local AdS$_4$ higher spin representation. One should note that the $1/N$ vertices do not become local in AdS spacetime. Actually the light-cone gauge fixing of Vasiliev's theory has not been established yet at the nonlinear level, so one would expect a nonlocal form.

Another important check regarding the identification of the ``extra'' AdS coordinate $z$ can be seen by taking the $z \rightarrow 0$ limit. Evaluating the bi-local field at $z=0$ gives the following ``boundary'' form
\begin{equation}
\text{$\mathcal{H}$}(x^+,x^-,x,\theta)=\int dp_1^+dp_2^+e^{ix^-(p_1^++p_2^+)}\delta(\theta-2\tan ^{-1}\sqrt{p_2^+/p_1^+})\tilde{\Psi}(p_1^+,p_2^+;x,x) \ .
\label{agreecurrent}
\end{equation}
Expanding the kernel in the above transformation into Fourier series, for a fixed even spin $s$, one has the binomial expansion
\begin{equation}
\left(\sqrt{p_1^+}-i\sqrt{p_2^+}\right)^{2s}=\frac{(-1)^k s! \; \Gamma(s+\frac{1}{2})\Gamma(\frac{1}{2})}{k!(s-k)! \; \Gamma(s-k+\frac{1}{2})\Gamma(k+\frac{1}{2})} (p_1^+)^k \, (p_2^+)^{s-k} \ .
\end{equation}
This is to be compared with conformal operators of a fixed spin $s$ which are explicitly given in \cite{Makeenko:1980bh, Braun:2003rp} by
\begin{equation}
\text{$\mathcal{O}$}^s=\sum_{k=0}^s\frac{(-1)^k \; \Gamma(s+\frac{1}{2})\Gamma(s+\frac{1}{2})}{k!(s-k)! \; \Gamma(s-k+\frac{1}{2})\Gamma(k+\frac{1}{2})}(\partial_+)^k \phi \, (\partial_+)^{s-k} \phi \ ,
\end{equation}
which agree with \eqref{agreecurrent} up to an overall normalization constant.

As a result, in the bi-local picture one has a clear definition of the boundary $z=0$ and the notion of boundary amplitudes (boundary $S$-matrix). Due to the construction through collective field theory, one is guaranteed to reproduce the boundary correlators in full agreement with the $O(N)$ model which are now reproduced in terms of Witten diagrams through higher $n$-point vertices as always in AdS duals. The bulk/bi-local theory is nonlinear with nonlinearities governed by $1/N=G_N$. All this provides a nontrivial check of the collective picture and the proposal that bi-local fields provide a bulk representation of AdS$_4$ higher spin fields.



\section{Coleman-Mandula Theorem in AdS$_\mathbf{4}$/CFT$_\mathbf{3}$} \label{sec:CM}

The simplest case of the correspondence involves the UV fixed point CFT of noninteracting $N$-component bosonic or fermionic fields. These theories are characterized by the existence of an infinite sequence of higher spin currents that are conserved. Consequently one has a higher symmetry and an infinite sequence of generators
\begin{equation}
Q^s=\int d\vec{x}J_{0\mu_1\mu_2\cdots \mu_s} \ .
\end{equation}

In such a theory, the Coleman-Mandula theorem would imply that the $S$-matrix should be 1. The relevance and implications of the Coleman-Mandula theorem in AdS$_4$/CFT$_3$ was recently addressed by Maldacena and Zhiboedov \cite{Maldacena:2011jn, Maldacena:2012sf}. They work in the light cone and make extensive use of only one particular charge
\begin{equation}
Q^s=\int dx^-dxJ_{---\cdots -} \ .
\end{equation}

In general there is a question regarding the existence of an $S$-matrix in CFT (and also AdS spacetime gravity). Maldacena and Zhiboedov considered the implication of the theorem on correlation functions; they demonstrated that the existence of the sequence of currents implies that the correlators are given by free fields, and it is in this sense that the theory can be categorized as simple, i.e. represented in terms of free fields. 

The recovered correlators $C_n=\langle \mathcal{O}_1 \mathcal{O}_2 \cdots \mathcal{O}_n \rangle$ even though expressible in terms of free $N$-component fields, are nonzero for all $n$. They are, as we have described above, associated with a nonlinear bulk theory, with nonlinearities governed by $1/N=G_N$. The question then concerns the fate of these nonlinearities characterizing the AdS$_4$ HS theory.

Boundary correlators are sometimes described in the literature as a ``boundary $S$-matrix'' of the AdS theory. In fact Mack \cite{Mack:2009mi, Mack:2009gy} has put forward arguments whereby CFT correlation functions themselves posses a structure equivalent to an $S$-matrix. He argued that they can be in general written in an integral form (the Mellin representation)
\begin{equation}
G(x_1 x_2\cdots x_3)=\int \prod ds_{ij} M(\{s_{ij}\}) \prod_{i<j} \Gamma(s_{ij})(x_{ij})^{-2 s_{ij}}
\end{equation}
which then implies various properties (crossing, duality, etc.) in support of their $S$-matrix interpretation. This interpretation was strengthened by the AdS calculation \cite{Penedones:2010ue}. Nevertheless, this ``boundary $S$-matrix'' lacks some of the key features of a genuine scattering matrix.

Based on the collective construction we would like to put forward (and evaluate) another more direct $S$-matrix which we will base on the physical picture of (collective) dipoles that underlie the CFT$_3$/Higher Spin Holography. In this picture we identify an appropriate on-shell relation and define the $S$-matrix through a Lehman-Symanzik-Zimmermann reduction formula. As such it will be given as a limit of general bi-local correlation functions.

\subsection{An example}

Before proceeding with the details, we describe an analogous example that features a simple (free) theory duality: the old $d=1$ Matrix Model / $2d$ non-critical string theory correspondence \cite{Das:1990kaa}. One has the matrix theory
\begin{equation}
L=\frac{1}{2}\text{Tr}\left(\dot{M}^2(t)+M^2(t)\right)
\end{equation}
and a Hamiltonian corresponding to $N^2$ decoupled harmonic oscillators
\begin{equation}
H=-\frac{1}{2}\sum_{\alpha=1}^{N^2}\left(\frac{\partial^2}{\partial M_\alpha^2}-M_\alpha^2\right) .
\end{equation}
In this model one also had an infinite sequence of higher charges: $Q_s=\text{Tr}[(P^2-M^2)^s]$ and an infinite $\mathcal{W}_{\infty}$ symmetry. In the basic matrix theory representation, there is clearly no scattering and no visible $S$-matrix. A spacetime interpretation of the model (and an $S$-matrix) is found through the collective (Fermi-Droplet) representation
\begin{equation}
M_{ij}(t)\rightarrow \phi(x,t)=\text{Tr} \, \delta(x-M_{ij}(t)) \ .
\end{equation}
The large $N$ collective Hamiltonian derived in \cite{Das:1990kaa} is given by
\begin{equation}
H_c=\int dx \left( \frac{1}{2} \partial_x \Pi(x) \phi(x) \partial_x \Pi(x)+\frac{\pi^2}{6}\phi^3-\frac{x^2}{2}\phi \right)
\end{equation}
where $\phi(x)$ and $\Pi(x)$ obey the canonical commutation relations $[\phi(x),\Pi(y)]=i\delta(x-y)$. This collective Hamiltonian correctly reproduces all the correlators $\langle \mathcal{O}_{n_1} \mathcal{O}_{n_2} \cdots \mathcal{O}_{n_k} \rangle$ for the most general invariant operators $\mathcal{O}_n=\text{Tr}(M^n)=\int dx \; x^n \phi \ .$

Small fluctuations of this (collective) theory $\phi=\phi_0+\partial_x \psi$, $\Pi=-\partial_x^{-1} \dot{\psi}$ features a 2d massless boson \cite{Demeterfi:1991cw}
\begin{equation}
H^{(2)}=\int d\sigma \left( \frac{1}{2}\dot{\psi}^2(t,\sigma)+\frac{\pi^2}{2}{\psi^\prime}^2(t,\sigma)\right) \ ,
\end{equation}
where the prime is the derivative with respect to the Liouville coordinate defined by
\begin{equation}
\sigma=\frac{1}{\pi} \int_0^x \frac{d y}{\phi_0(y)} \ .
\end{equation}
Consequently one is led to consider the scattering of collective massless bosons \cite{Gross:1991qp} with an on-shell condition: $K_\mu=(E,K)$ and $E^2-K^2=0$. Evaluation of the corresponding scattering amplitudes gives the $S$-matrix. For the three-point scattering amplitude, one has
\begin{eqnarray}
S_3(E_1,E_2,E_3) &=& 2\pi\delta(E_1+E_2+E_3) \left[ \prod_{i=1}^3 (E_i-K_i)- \prod_{i=1}^3 (E_i+K_i) \right] \cr
&=& 2\pi\delta(E_1+E_2+E_3) \left[ \prod_{i=1}^3 \bigl( E_i-\vert E_i\vert \bigr)- \prod_{i=1}^3 \bigl(E_i+\vert E_i\vert \bigr) \right]
\end{eqnarray}
where we have used $K_i=\vert E_i\vert$ (corresponding to Liouville as time). For the scattering of incoming (outgoing) particles, we have $E_1,E_2>0$, $E_3<0$, so that
\begin{eqnarray}
S_3(+,+,-)=0 \ .
\end{eqnarray}
In the same way one can show $S_{n \ge 4}=0$. A change of boundary conditions (in particular Dirichlet), gives however a non-trivial result $S_n \neq 0$ which was then compared with the string scattering amplitudes.

\subsection{Evaluation of the three- and four-point amplitudes}

Let us now return to the bi-local theory and consider therefore the $S$-matrix for scattering of ``collective dipoles''. In a time-like gauge (single-time), one has the on-shell relation: $E^2-(\vert \vec{k}_1\vert+\vert \vec{k}_2\vert)^2=0$, and the $S$-matrix can be defined by the LSZ-type reduction formula
\begin{eqnarray}
S=\lim \prod_{i} (E_i^2-(\vert \vec{k}_i\vert+\vert \vec{k}_{i'}\vert )^2) \langle \tilde{\Psi}(E_1,\vec{k}_1,\vec{k}_{1'})\tilde{\Psi}(E_2,\vec{k}_2,\vec{k}_{2'})\cdots \rangle
\label{Sdefintion}
\end{eqnarray}
where the $\tilde{\Psi}$ operators denote energy-momentum transforms of the bi-local fields \eqref{bilocal}. The limit implies the on-shell specification for the energies of the dipoles.
In the light-cone gauge, (\ref{Sdefintion}) would correspond to
\begin{eqnarray}
\lim \prod_{i} (P_i^--\frac{p_i^2}{2p_i^+}-\frac{p^2_{i'}}{2p^+_{i'}}) \langle \tilde{\Psi}(P_1^-;p_1^+,p_1,p^+_{1'},p_{1'}) \tilde{\Psi} (P_2^-;p^+_2,p_2,p^+_{2'},p_{2'})\cdots \rangle \ .
\end{eqnarray}
We note that the correlation functions appearing in this construction are not the correlation functions of conformal current operators $J_{--\cdots-}$. As Maldacena and Zhiboedov have discussed, the Ward identities based on  currents provide a reconstruction of correlation functions for bi-local operators of the form $\mathcal{B}(x^+; (x_1^-,x_2^-); x_1=x_2)$. Since these are bi-local in $x$ but local in the other coordinates one is not in a position to consider the above defined $S$-matrix.

Our evaluation of the $S$-matrix proceeds as follows. Using the time-like quantization we will evaluate the 3 and 4-point scattering amplitude corresponding to associated Witten diagrams. In momentum space, in terms of the bi-local fields
\begin{eqnarray}
\eta(t;\vec{x}_1,\vec{x}_2) &=& \int d\vec{k}_1 d\vec{k}_2\frac{1}{\sqrt{2\omega_{k_1}\omega_{k_2}}} \left(e^{+i(\vec{k}_1 \cdot \vec{x}_1+\vec{k}_2 \cdot \vec{x}_2)}\alpha_{\vec{k}_1\vec{k}_2}+h.c. \right) \\
\pi(t;\vec{x}_1,\vec{x}_2) &=& i \int d\vec{k}_1 d\vec{k}_2 \sqrt{\frac{\omega_{k_1}\omega_{k_2}}{2}} \left(e^{-i(\vec{k}_1 \cdot \vec{x}_1+\vec{k}_2 \cdot \vec{x}_2)}\alpha^\dagger_{\vec{k}_1\vec{k}_2}-h.c. \right)
\end{eqnarray}
the cubic (\ref{cubic}) and quartic (\ref{quartic}) interaction potentials take the form
\begin{eqnarray}
H^{(3)} &=& \frac{\sqrt{2}}{\sqrt{N}}\int \prod_{i=1}^3 d\vec{k}_i \Bigl[-\frac{\omega_{k_1 k_2 k_3}}{3}\alpha_{\vec{k}_1 \vec{k}_2}\alpha_{-\vec{k}_2 \vec{k}_3}\alpha_{-\vec{k}_3-\vec{k}_1}+\omega_{k_2} \alpha_{\vec{k}_1 \vec{k}_2}\alpha_{-\vec{k}_2 \vec{k}_3}\alpha^\dagger_{\vec{k}_3 \vec{k}_1}+h.c. \Bigr] \label{cubicosc} \\
H^{(4)} &=& \frac{1}{N} \int \prod_{i=1}^4 d\vec{k}_i \; \frac{\omega_{k_1 k_2 k_3 k_4}}{4} \Bigl[\alpha_{\vec{k}_1 \vec{k}_2}\alpha_{-\vec{k}_2 \vec{k}_3}\alpha_{-\vec{k}_3 \vec{k}_4}\alpha_{-\vec{k}_4-\vec{k}_1}+4\alpha_{\vec{k}_1 \vec{k}_2}\alpha_{-\vec{k}_2 \vec{k}_3}\alpha_{-\vec{k}_3 \vec{k}_4}\alpha^\dagger_{\vec{k}_4 \vec{k}_1}+h.c. \cr
&& \qquad\qquad\qquad\qquad+4\alpha_{\vec{k}_1 \vec{k}_2}\alpha_{-\vec{k}_2 \vec{k}_3}\alpha^\dagger_{\vec{k}_3 \vec{k}_4}\alpha^\dagger_{-\vec{k}_4 \vec{k}_1}+2\alpha_{\vec{k}_1 \vec{k}_2}\alpha^\dagger_{\vec{k}_2 \vec{k}_3}\alpha_{\vec{k}_3 \vec{k}_4}\alpha^\dagger_{\vec{k}_4 \vec{k}_1} \Bigr]
\label{quarticosc}
\end{eqnarray}
where we used the notation $\omega_{k_1k_2\cdots k_i} \equiv \omega_{k_1}+\omega_{k_2}+\cdots+\omega_{k_i}$ and $h.c.$ means taking the hermitian conjugate of {\it only} the terms ahead of it.

\begin{figure}
\begin{center}
\includegraphics[width=0.3\textwidth]{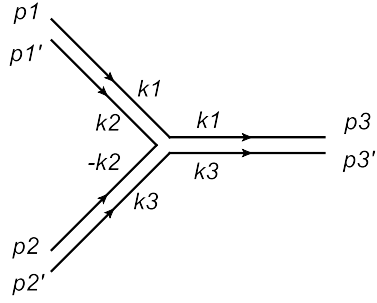}  
\caption{Scattering of three dipoles.}
\label{threescattering}
\end{center}
\end{figure}

For the three-dipole scattering ($1+2 \to 3$), the amplitude is given by
\begin{equation}
\langle 0 \vert \alpha_{\vec{p}_{3} \vec{p}_{3'}} \,
T \exp \left[ -i \int_{-\infty}^\infty dt \, H^{(3)} (t) \right]
\alpha^\dagger_{\vec{p}_{2} \vec{p}_{2'}} \alpha^\dagger_{\vec{p}_{1} \vec{p}_{1'}} \vert 0 \rangle
\end{equation}
where $T$ means {\it time-ordered}. Using the explicit form of the cubic interaction given in (\ref{cubicosc}), the only surviving $1/\sqrt{N}$ contribution to the scattering amplitude is
\begin{equation}
-\frac{i\sqrt{2}}{\sqrt{N}} \int d\vec{k}_i \; \omega_{k_2}
\langle 0 \vert \alpha_{\vec{p}_{3} \vec{p}_{3'}}
\alpha_{\vec{k}_1 \vec{k}_2} \alpha_{-\vec{k}_2 \vec{k}_3} \alpha^\dagger_{\vec{k}_3 \vec{k}_1}
\alpha^\dagger_{\vec{p}_{2} \vec{p}_{2'}} \alpha^\dagger_{\vec{p}_{1} \vec{p}_{1'}} \vert 0 \rangle \ .
\label{vev3}
\end{equation}
Graphically, this corresponds to evaluate the Feynman diagram shown in Figure \ref{threescattering}.

The evaluation can be most easily performed in the {\it interaction picture}. Using the bi-local propagator symmetrized over the momenta
\begin{align}
\langle 0\vert T \alpha_{\vec{p}_{1} \vec{p}_{1'}}(t_1)
\alpha^\dagger_{\vec{p}_{2} \vec{p}_{2'}}(t_2)\vert 0 \rangle
=\int dE\frac{ie^{-iE(t_1-t_2)}}{E-\omega_{p_1}-\omega_{p_{1'}}}
& \frac{1}{2}[\delta(\vec{p}_1-\vec{p}_2)\delta(\vec{p}_{1'}-\vec{p}_{2'}) \cr
& +\delta(\vec{p}_1-\vec{p}_{2'})\delta(\vec{p}_{1'}-\vec{p}_2)] \ ,
\end{align}
the integral over the vev in (\ref{vev3}) is calculated to be
\begin{align}
& \frac{i^3}{8} \delta(E_1+E_2-E_3) [(\omega_{p_{1'}}+\omega_{p_2}) \delta(\vec{p}_1-\vec{p}_3) \delta( \vec{p}_{2'}-\vec{p}_{3'}) \delta (\vec{p}_{1'}+\vec{p}_2) + \text{7 more terms} ] \cr
=& \frac{i^3}{8} (E_1+E_2-E_3) \delta(E_1+E_2-E_3) [\delta(\vec{p}_1-\vec{p}_3) \delta( \vec{p}_{2'}-\vec{p}_{3'}) \delta (\vec{p}_{1'}+\vec{p}_2) + \cdots ]
\end{align}
where in the last step we have used energy conservation and the delta functions. The seven more terms are due to the symmetrization of $(1 \leftrightarrow 1'),(2 \leftrightarrow 2'),(3 \leftrightarrow 3')$. Combining all the pre-factors, we get the final result 
\begin{align}
S(1+2\rightarrow 3)=&-\frac{\sqrt{2}}{8\sqrt{N}} (E_1+E_2-E_3) \; \delta(E_1+E_2-E_3) \cr
&\{\delta(\vec{p}_1-\vec{p}_3)\delta(\vec{p}_{2'}-\vec{p}_{3'})\delta(\vec{p}_{1'}+\vec{p}_2)+\text{7 more terms} \}
\end{align}
so that the result $S_3=0$ follows.

\begin{figure}
\begin{center}
\includegraphics[width=0.35\textwidth]{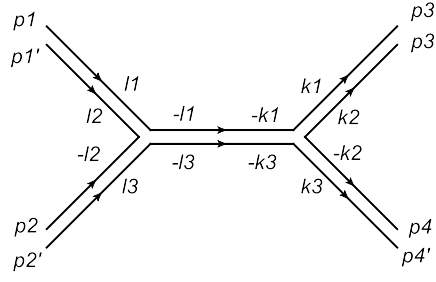}  
\caption{The $s$-channel diagram of four-dipole scattering.}
\label{fourchanneelsca}
\end{center}
\end{figure}

Next for the four-dipole scattering ($1+2 \to 3+4$), the calculation is similar. The scattering amplitude is given by
\begin{equation}
\langle 0\vert \alpha_{\vec{p}_{3}\vec{p}_{3'}}\alpha_{\vec{p}_{4}\vec{p}_{4'}} T \exp \left[
-i\int^\infty_{-\infty} dt \, \left( H^{(3)} (t) + H^{(4)} (t) \right) \right]
\alpha^\dagger_{\vec{p}_{1} \vec{p}_{1'}}\alpha^\dagger_{\vec{p}_{2} \vec{p}_{2'}}\vert 0 \rangle
\end{equation}
where $H^{(4)}$ is explicitly given in (\ref{quarticosc}). The $1/N$ contributions to the $S_4$ scattering amplitude are collected as follows
\begin{eqnarray}
&&-\frac{2}{9N}\int  d\vec{k}_i d\vec{l}_j \, \omega_{k_1k_2k_3}\omega_{l_1l_2l_3}\langle 0 \vert \alpha_{\vec{p}_{3} \vec{p}_{3'}}\alpha_{\vec{p}_{4} \vec{p}_{4'}}\alpha_{\vec{k}_1 \vec{k}_2}\alpha_{-\vec{k}_2 \vec{k}_3}\alpha_{-\vec{k}_3-\vec{k}_1}\alpha^\dagger_{\vec{l}_1 \vec{l}_2}\alpha^\dagger_{-\vec{l}_2 \vec{l}_3}\alpha^\dagger_{-\vec{l}_3-\vec{l}_1}\alpha^\dagger_{\vec{p}_{1} \vec{p}_{1'}}\alpha^\dagger_{\vec{p}_{2} \vec{p}_{2'}}\vert 0 \rangle \cr
&&-\frac{2}{N}\int  d\vec{k}_i d\vec{l}_j \, \omega_{k_2}\omega_{l_2}\langle 0 \vert \alpha_{\vec{p}_{3} \vec{p}_{3'}}\alpha_{\vec{p}_{4} \vec{p}_{4'}}\alpha_{\vec{k}_1 \vec{k}_2}\alpha_{-\vec{k}_2 \vec{k}_3}\alpha^\dagger_{\vec{k}_3 \vec{k}_1}\alpha^\dagger_{\vec{l}_1 \vec{l}_2}\alpha^\dagger_{-\vec{l}_2 \vec{l}_3}\alpha_{\vec{l}_3 \vec{l}_1}\alpha^\dagger_{\vec{p}_{1} \vec{p}_{1'}}\alpha^\dagger_{\vec{p}_{2} \vec{p}_{2'}}\vert 0 \rangle \cr
&&-\frac{i}{N}\int  d\vec{k}_i \, \omega_{k_1k_2k_3k_4}\langle 0\vert \alpha_{\vec{p}_{3} \vec{p}_{3'}}\alpha_{\vec{p}_{4} \vec{p}_{4'}}\alpha_{\vec{k}_1 \vec{k}_2}\alpha_{-\vec{k}_2 \vec{k}_3}\alpha^\dagger_{\vec{k}_3 \vec{k}_4}\alpha^\dagger_{-\vec{k}_4 \vec{k}_1}\alpha^\dagger_{\vec{p}_{1} \vec{p}_{1'}}\alpha^\dagger_{\vec{p}_{2} \vec{p}_{2'}}\vert 0 \rangle \cr
&&-\frac{i}{2N}\int  d\vec{k}_i \, \omega_{k_1k_2k_3k_4}\langle 0 \vert \alpha_{\vec{p}_{3} \vec{p}_{3'}}\alpha_{\vec{p}_{4} \vec{p}_{4'}}\alpha_{\vec{k}_1 \vec{k}_2}\alpha^\dagger_{\vec{k}_2 \vec{k}_3}\alpha_{\vec{k}_3 \vec{k}_4}\alpha^\dagger_{\vec{k}_4 \vec{k}_1}\alpha^\dagger_{\vec{p}_{1} \vec{p}_{1'}}\alpha^\dagger_{\vec{p}_{2} \vec{p}_{2'}}\vert 0 \rangle \ .
\label{fourlines}
\end{eqnarray}
The first line of (\ref{fourlines}) has only $s$-channel contributions shown in Figure \ref{fourchanneelsca}, while the second line of (\ref{fourlines}) has all $s,t,u$-channel contributions. The $s$-channel diagrams and their twisted ones (due to the symmetrization of propagators) are summed to be
\begin{align}
\frac{i}{8 N} \delta(E_1+E_2-E_3-E_4) \bigl[ & (\omega_{p_{2'}}+\omega_{p_3}) \delta(\vec{p}_1-\vec{p}_3) \delta(\vec{p}_{1'}+\vec{p}_2) \cr
& \delta( \vec{p}_{2'}-\vec{p}_{4'}) \delta (\vec{p}_{3'}+\vec{p}_4) + \text{15 more terms} \cr
+ & (\omega_{p_{1'}}+\omega_{p_{3}}) \delta(\vec{p}_2-\vec{p}_3) \delta(\vec{p}_1+\vec{p}_{2'}) \cr
& \delta( \vec{p}_{1'}-\vec{p}_{4'}) \delta (\vec{p}_{3'}+\vec{p}_4) + \text{15 more terms} \bigr] \ .
\end{align}

\noindent It is also convenient to calculate the $t,u$-channel diagrams together, with their twisted diagrams, they are summed to be
\begin{align}
\frac{i}{8 N} \delta(E_1+E_2-E_3-E_4) \bigl[ & (\omega_{p_{1'}}+\omega_{p_2}) \delta(\vec{p}_1-\vec{p}_3) \delta(\vec{p}_{1'}+\vec{p}_2) \cr
& \delta( \vec{p}_{2'}-\vec{p}_{4'}) \delta (\vec{p}_{3'}+\vec{p}_4) + \text{15 more terms} \cr
+ & (\omega_{p_1}+\omega_{p_{2'}}) \delta(\vec{p}_2-\vec{p}_3) \delta(\vec{p}_1+\vec{p}_{2'}) \cr
& \delta( \vec{p}_{1'}-\vec{p}_{4'}) \delta (\vec{p}_{3'}+\vec{p}_4) + \text{15 more terms} \bigr] \\
+ \frac{i}{16 N} \delta(E_1+E_2-E_3-E_4) \bigl[ & (\omega_{p_{1}}+\omega_{p_{1'}}+\omega_{p_2}+\omega_{p_{2'}}) \delta(\vec{p}_1-\vec{p}_3) \delta(\vec{p}_{1'}-\vec{p}_4) \cr
& \delta( \vec{p}_{2'}-\vec{p}_{4'}) \delta (\vec{p}_2-\vec{p}_{3'}) + \text{15 more terms} \cr
+ & (\omega_{p_{1}}+\omega_{p_{1'}}+\omega_{p_2}+\omega_{p_{2'}}) \delta(\vec{p}_2-\vec{p}_3) \delta(\vec{p}_{2'}-\vec{p}_4) \cr
& \delta( \vec{p}_{1'}-\vec{p}_{4'}) \delta (\vec{p}_1-\vec{p}_{3'}) + \text{15 more terms} \bigr] \ .
\label{cancel1}
\end{align}

\begin{figure}
\begin{center}
\includegraphics[width=0.3\textwidth]{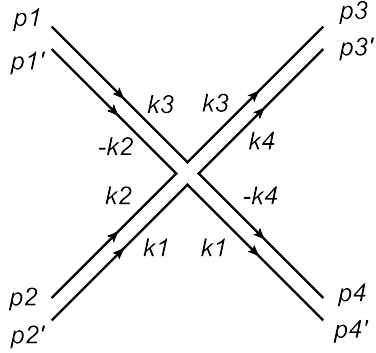}  
\caption{The cross-shaped diagram of four-dipole scattering.}
\label{fourcrosssca}
\end{center}
\end{figure}

\noindent The third line of (\ref{fourlines}) is the cross-shaped diagram shown in Figure \ref{fourcrosssca}, which gives the result
\begin{align}
 -\frac{i}{8 N} \delta(E_1+E_2-E_3-E_4) \bigl[ & (\omega_{p_{1'}}+\omega_{p_{2'}}+\omega_{p_3}+\omega_{p_4}) \delta(\vec{p}_1-\vec{p}_3) \delta(\vec{p}_{1'}+\vec{p}_2) \cr
& \delta( \vec{p}_{2'}-\vec{p}_{4'}) \delta (\vec{p}_{3'}+\vec{p}_4) + \text{15 more terms} \cr
+ & (\omega_{p_{1}}+\omega_{p_{2}}+\omega_{p_{3'}}+\omega_{p_{4'}}) \delta(\vec{p}_2-\vec{p}_3) \delta(\vec{p}_1+\vec{p}_{2'}) \cr
& \delta( \vec{p}_{1'}-\vec{p}_{4'}) \delta (\vec{p}_{3'}+\vec{p}_4) + \text{15 more terms} \bigr] \ .
\end{align}

\noindent The calculation of the fourth line is similar to the third line, which gives the result
\begin{align}
 -\frac{i}{16 N} \delta(E_1+E_2-E_3-E_4) \bigl[ & (\omega_{p_{1}}+\omega_{p_{1'}}+\omega_{p_2}+\omega_{p_{2'}}) \delta(\vec{p}_1-\vec{p}_3) \delta(\vec{p}_{1'}-\vec{p}_4) \cr
& \delta( \vec{p}_{2'}-\vec{p}_{4'}) \delta (\vec{p}_2-\vec{p}_{3'}) + \text{15 more terms} \cr
+ & (\omega_{p_{1}}+\omega_{p_{1'}}+\omega_{p_2}+\omega_{p_{2'}}) \delta(\vec{p}_2-\vec{p}_3) \delta(\vec{p}_{2'}-\vec{p}_4) \cr
& \delta( \vec{p}_{1'}-\vec{p}_{4'}) \delta (\vec{p}_1-\vec{p}_{3'}) + \text{15 more terms} \bigr] \ .
\label{cancel2}
\end{align}

\noindent Summing all the diagrams, it is easy to see (\ref{cancel1}) and (\ref{cancel2}) cancel each other, while the rest diagrams give the final result
\begin{align}
 S(1+ & 2\rightarrow 3+4)= \frac{i}{8 N} \delta(E_1+E_2-E_3-E_4) \cr
& \bigl[ (\omega_{p_2}-\omega_{p_4}) \delta(\vec{p}_1-\vec{p}_3) \delta(\vec{p}_{1'}+\vec{p}_2) \delta( \vec{p}_{2'}-\vec{p}_{4'}) \delta (\vec{p}_{3'}+\vec{p}_4) + \text{15 more terms} \cr
+ & (\omega_{p_{2'}}-\omega_{p_{3'}}) \delta(\vec{p}_2-\vec{p}_3) \delta(\vec{p}_1+\vec{p}_{2'}) \delta( \vec{p}_{1'}-\vec{p}_{4'}) \delta (\vec{p}_{3'}+\vec{p}_4) + \text{15 more terms} \bigr] \cr
=& \frac{i}{16 N} (E_1+E_2-E_3-E_4) \delta(E_1+E_2-E_3-E_4)  \cr
& \bigl[ \delta(\vec{p}_1-\vec{p}_3) \delta(\vec{p}_{1'}+\vec{p}_2) \delta( \vec{p}_{2'}-\vec{p}_{4'}) \delta (\vec{p}_{3'}+\vec{p}_4) + \text{15 more terms} \cr
+ & \; \delta(\vec{p}_2-\vec{p}_3) \delta(\vec{p}_1+\vec{p}_{2'}) \delta( \vec{p}_{1'}-\vec{p}_{4'}) \delta (\vec{p}_{3'}+\vec{p}_4) + \text{15 more terms} \bigr] \ ,
\end{align}
which implies $S_4=0$.


It is clear that the direct evaluation can be continued to higher point scattering with the conjectured result $S_{n \ge 5}=0$. One can describe the nonlinear collective field theory in the following way: its nonlinearity, and higher point vertices are precisely such that they reproduce the boundary correlators through bi-local (Witten) diagrams. These same diagrams however give vanishing results in the on-shell evaluation as described above.

In general quantum field theory, one has the equivalence theorem. Consequently a vanishing $S$-matrix implies that there should exist a (nonlinear) field transformation which linearizes the theory. For the present case this concerns the linearization of bulk $G_N=1/N$ interactions. We will in the next section describe such a field transformation. 

Since we view the collective construction to represent a gauge fixed description of Vasiliev's HS theory, analogous statements are expected to hold there. Finally it is also clear that one can expect that any change of boundary conditions will result in non-trivial $S$-matrix.



\section{Field Transformation} \label{sec:Field}

We have concluded in the previous section that the $S$-matrix equals $1$ for the bi-local theory of the free UV fixed point. The theory is nonlinear with a sequence of $1/N$ vertices which are needed to reproduce arbitrary $n$-point correlators (and the ``boundary $S$-matrix''). By correspondence Vasiliev's HS theory has the same properties. As suggested in section \ref{sec:CM}, this implies that there should be a field transformation that linearizes the $G_N=1/N$ interactions. We will now describe such a procedure for deducing the transformation. The procedure is based on considering an algebraic description of the bi-local system. We will be able to show that the bi-local pseudo-spin algebra has among other two representations: one equalling the nonlinear collective field theory and another in which the Hamiltonian becomes quadratic.


For the free theory in question one has exact creation operators for the singlet sector of the theory. They are given by the bi-local operators
\begin{eqnarray}
A(\vec{p}_1,\vec{p}_2)&=&\frac{1}{\sqrt{2N}}\sum_i a^i(\vec{p}_1)a^i(\vec{p}_2) \ , \\
A^\dagger(\vec{p}_1,\vec{p}_2)&=&\frac{1}{\sqrt{2N}}\sum_i a^{i\, \dagger}(\vec{p}_1)a^{i\, \dagger}(\vec{p}_2) \ , \\
B(\vec{p}_1,\vec{p}_2)&=&\frac{1}{2}\sum_i a^{i\,\dagger}(\vec{p}_1)a^i(\vec{p}_2) \ .
\end{eqnarray}
In terms of these collective variables the Hamiltonian is
\begin{equation}
H=\int d^{d-1}\vec{p} \; {\cal H}(\vec{p},\vec{p}) \ , \qquad
{\cal H}(\vec{p},\vec{p}) = 2\omega_{\vec{p}} \, B(\vec{p},\vec{p})
+ \frac{N}{2}\omega_{\vec{p}} \, \delta(\vec{0}) \ .
\label{Hamiltonian}
\end{equation}
The above operators (representing bi-local pseudo-spin variables) close an algebra
\begin{align}
\big[ A(\vec{p}_1,\vec{p}_2),A^\dagger(\vec{p}_3,\vec{p}_4)\big]=
& \, \frac{1}{2}\left(\delta_{\vec{p}_2,\vec{p}_3}\delta_{\vec{p}_4,\vec{p}_1}+\delta_{\vec{p}_2,\vec{p}_4}\delta_{\vec{p}_3,\vec{p}_1}\right)
+\frac{1}{N}\bigl[\delta_{\vec{p}_2,\vec{p}_3}B(\vec{p}_4,\vec{p}_1) \cr
&+\delta_{\vec{p}_2,\vec{p}_4}B(\vec{p}_3,\vec{p}_1)
+\delta_{\vec{p}_1,\vec{p}_3}B(\vec{p}_4,\vec{p}_2)
+\delta_{\vec{p}_1,\vec{p}_4}B(\vec{p}_3,\vec{p}_2) \bigr] \ , \\
\big[ B(\vec{p}_1,\vec{p}_2),A^\dagger(\vec{p}_3,\vec{p}_4)\big]=
& \, \frac{1}{2}\bigl( \delta_{\vec{p}_2,\vec{p}_3}A^\dagger(\vec{p}_1,\vec{p}_4)+
\delta_{\vec{p}_2,\vec{p}_4}A^\dagger(\vec{p}_1,\vec{p}_3)\bigr) \ , \\
\big[ B(\vec{p}_1,\vec{p}_2),A(\vec{p}_3,\vec{p}_4)\big]=
&-\frac{1}{2}\bigl(\delta_{\vec{p}_1,\vec{p}_3}A(\vec{p}_2,\vec{p}_4)+
\delta_{\vec{p}_1,\vec{p}_4}A(\vec{p}_2,\vec{p}_3)\bigr) \ .
\end{align}

We note that the theory based on this algebra was studied in detail by Berezin \cite{Berezin:1978sn}. In the $O(N)$ case one finds the quadratic (Casimir) constraint
\begin{equation}
-\frac{8}{N}A^\dagger\star A +\left(1+\frac{4}{N}B\right)\star \left(1+\frac{4}{N}B\right) = \text{$\mathbb{I}$} \ .
\end{equation}
The importance of the Casimir constraint is that it implies that the above non-commuting set of bi-local operators is not independent. In particular the bi-local pseudo-spin algebra has representations in terms of canonical pairs of variables.

The canonical collective theory based on the equal-time bi-local field and its conjugate provides one specific representation of the above algebra. Explicitly, one can show
\begin{eqnarray}
A(\vec{x}_1,\vec{x_2})&=&\int d\vec{p}_1d\vec{p}_2d\vec{y}_1d\vec{y}_2e^{i\vec{p}_1\cdot(\vec{x}_1-\vec{y}_1)}e^{i\vec{p}_2\cdot(\vec{x}_2-\vec{y}_2)}
\Bigl[\frac{-2}{\sqrt{\omega_{p_1} \omega_{p_2}}}\Pi(\vec{y}_1,\vec{z}_1)\star\Psi(\vec{z}_1,\vec{z}_2)\star\Pi(\vec{z}_2,\vec{y}_2)\cr
&&-i\sqrt{N}\sqrt{\frac{\omega_{p_2}}{\omega_{p_1}}}\Psi(\vec{y}_2,\vec{z}_1)\star\Pi(\vec{y}_1,\vec{z}_1)
-i\sqrt{N}\sqrt{\frac{\omega_{p_1}}{\omega_{p_2}}}\Psi(\vec{y}_1,\vec{z}_1)\star\Pi(\vec{y}_2,\vec{z}_1)\cr
&&-\frac{N}{8}\frac{1}{\sqrt{\omega_{p_1} \omega_{p_2}}}\frac{1}{\Psi}(\vec{y}_1,\vec{y}_2)
+\frac{N\sqrt{\omega_{p_1} \omega_{p_2}}}{2}\Psi(\vec{y}_1,\vec{y}_2) \Bigr] \ .
\end{eqnarray}
Transforming it to momentum space and expanding in $1/N$ we generate an infinite series
\begin{align}
A(\vec{k}_1,\vec{k}_2)=& \, \alpha_{\vec{k}_1\vec{k}_2}-\frac{1}{\sqrt{2N}}\Bigl[
\alpha_{\vec{k}_1\vec{k}_3}\star\alpha_{-\vec{k}_3\vec{k}_2}
-\alpha_{\vec{k}_1\vec{k}_3}^\dagger\star\alpha_{-\vec{k}_3\vec{k}_2}^\dagger \cr
& -\alpha_{\vec{k}_1\vec{k}_3}\star\alpha_{\vec{k}_3\vec{k}_2}^\dagger
-\alpha_{\vec{k}_1\vec{k}_3}\star\alpha_{\vec{k}_3\vec{k}_2}^\dagger\Bigr]+O(\alpha^3) \ , \\
B(\vec{k}_1,\vec{k}_2)=& \, \frac{1}{2}\left[
\alpha_{\vec{k}_1\vec{k}_3}\star\alpha_{\vec{k}_3\vec{k}_2}^\dagger
+\alpha_{\vec{k}_1\vec{k}_3}^\dagger\star\alpha_{\vec{k}_3\vec{k}_2}\right]
+\sqrt{\frac{2}{N}}\Bigl[\alpha_{\vec{k}_1\vec{k}_3}\star\alpha_{-\vec{k}_3\vec{k}_4}
\star\alpha_{-\vec{k}_4\vec{k}_2} \cr
&+\alpha_{\vec{k}_1\vec{k}_3} \star\alpha_{\vec{k}_3\vec{k}_4}^\dagger\star\alpha_{\vec{k}_4\vec{k}_2}
-\alpha_{\vec{k}_1\vec{k}_3}^\dagger\star\alpha_{\vec{k}_3\vec{k}_4}\star\alpha_{\vec{k}_4\vec{k}_2}^\dagger \cr
&-\alpha_{\vec{k}_1\vec{k}_3}^\dagger\star\alpha_{-\vec{k}_3\vec{k}_4}^\dagger
\star\alpha_{-\vec{k}_4\vec{k}_2}^\dagger \Bigr]+O(\alpha^4) \ .
\end{align}
The key to our arguments is the fact that one can write another realization of the algebra in terms of an oscillator $\beta (\vec{p}_1,\vec{p}_2)$ obeying
\begin{eqnarray}
\beta(\vec{p}_1,\vec{p}_2) &=& \left(1+\frac{2}{N} B\right)^{-\frac{1}{2}}(\vec{p}_1,\vec{p})
\star A(\vec{p},\vec{p}_2) \label{invbetabilocalrep} \\
\beta^\dagger(\vec{p}_1,\vec{p}_2) &=& A^\dagger(\vec{p}_1,\vec{p})
\star \left(1+\frac{2}{N}B \right)^{-\frac{1}{2}}(\vec{p},\vec{p}_2)
\end{eqnarray}
which has two important properties that
\begin{eqnarray}
B(\vec{p}_1,\vec{p}_2) &=& \beta^\dagger(\vec{p}_1,\vec{p})\star \beta(\vec{p},\vec{p}_2) \\
\left[\beta (\vec{p}_1,\vec{p}_2),\beta^\dagger (\vec{p}_3,\vec{p}_4)\right] &=& \delta_{\vec{p}_1,\vec{p}_4}\delta_{\vec{p}_2,\vec{p}_3} \ .
\end{eqnarray}
We see that in this realization the Hamiltonian is quadratic due to \eqref{Hamiltonian}. Furthermore, using \eqref{invbetabilocalrep} one can generate the transformation between the fields
\begin{align}
\beta(\vec{k}_1,\vec{k}_2)=& \, \alpha_{\vec{k}_1\vec{k}_2}-\frac{1}{\sqrt{2N}}\Bigl[
\alpha_{\vec{k}_1\vec{k}_3}\star\alpha_{-\vec{k}_3\vec{k}_2}
-\alpha_{\vec{k}_1\vec{k}_3}^\dagger\star\alpha_{-\vec{k}_3\vec{k}_2}^\dagger \cr
&-\alpha_{\vec{k}_1\vec{k}_3}\star\alpha_{\vec{k}_3\vec{k}_2}^\dagger
-\alpha_{\vec{k}_1\vec{k}_3}\star\alpha_{\vec{k}_3\vec{k}_2}^\dagger \Bigr]+O(\alpha^3) \ .
\end{align}

In conclusion we have presented a construction of the field transformation (in bi-local space) that linearizes the nonlinear $1/N$ Hamiltonian. Under this transformation the correlation functions change but the $S$-matrix does not. This represents the working of the Coleman-Mandula theorem in the large $N$ dual associated with the free field CFT. As such it complements the Maldacena-Zhiboedov argument for these theories.



\section{Conclusion} \label{sec:Con}

We have discussed some features of the Higher Spin AdS correspondence involving free $O(N)$ fields. The existence of an (infinite) sequence of higher symmetries in these theories raises the question regarding the implementation of the Coleman-Mandula theorem. Our focus was the question regarding the nonlinear $1/N$ theory which reproduces the (boundary) correlators. We argued that in these theories we are able to define a genuine $S$-matrix representing the scattering of collective dipoles. The $S$-matrix is specified with the standard LSZ procedure as an on-shell limit of (bi-local) correlation functions. 

For the theory based on the free correspondence i.e. the UV fixed point of the vector model we have evaluated the $S$-matrix showing the result $S=1$ as claimed in the title. This represents the consequence of the Coleman-Mandula theorem for the associated Higher Spin theory and complements the results of Maldacena and Zhiboedov. As we have discussed it implies that the nonlinear Higher Spin theory can be linearized through nonlinear field transformations. We have explicitly constructed such a transformation in the bi-local framework. We have also emphasized that a change of boundary conditions will change the above conclusion, namely one expects a nontrivial $S$-matrix. Based on the present results and the earlier $c=1$ case it is plausible to conclude that these features will characterize any large $N$ correspondence based on free fields.



\section*{Acknowledgments}

We would like to thank Sumit Das, Igor Klebanov, Soo-Jong Rey, Juan Maldacena and Suvrat Raju for relevant and constructive comments. The work of AJ and QY is supported by the Department of Energy under contract DE-FG-02-91ER40688. RdMK is supported by the South African Research Chairs Initiative of the Department of Science and Technology and National Research Foundation. The work of KJ is supported in part by the Swiss National Science Foundation. 

Some of the results were presented by AJ at ``CQUeST Spring Workshop on Higher Spins and String Geometry, Seoul'' and ``ESI Workshop on Higher Spin Gravity, Vienna''. He would like to thank the organizers for their hospitality. KJ would also like to thank the Erwin Schr\"{o}dinger Institute, Vienna.







\end{document}